\documentclass[a4paper,
               biblatex,     % biblatex is used
               keeplastbox,   % flushend option: not to un-indent last line in References
               ]{jacow}
\usepackage{pdfpages,multirow,ragged2e} %
\makeatletter%
	\ifboolexpr{bool{xetex}}
	 {\renewcommand{\Gin@extensions}{.pdf,%
	                    .png,.jpg,.bmp,.pict,.tif,.psd,.mac,.sga,.tga,.gif,%
	                    .eps,.ps,%
	                    }}{}
\makeatother

\ifboolexpr{bool{xetex} or bool{luatex}} % test for XeTeX/LuaTeX
 {}                                      % input encoding is utf8 by default
 {\usepackage[utf8]{inputenc}}           % switch to utf8

\usepackage[USenglish]{babel}

\ifboolexpr{bool{jacowbiblatex}}%
 {%
  \addbibresource{MOPL152.bib}
 }{}
\begin{document}

\title{Positron Beams at \NoCaseChange{Ce\textsuperscript{+}BAF}\thanks{This project is supported by the U.S. Department of Energy, Office of Science, Office of Nuclear Physics under contract DE-AC05-06OR23177; UT-Battelle, LLC, under contract DE-AC05-00OR22725 with the US Department of Energy (DOE); the European Union's Horizon 2020 research and innovation program under agreement STRONG - 2020 – No. 824093; the Programa de Fomento y Apoyo a Proyectos de Investigación code A1-022, from the Universidad Autónoma de Sinaloa.}}

\author{
J.~Grames$^1$,  %\thanks{grames@jlab.org} 
J.~Benesch$^1$,
M.~Bruker$^1$,  
L.~Cardman$^1$,
S.~Covrig$^1$, 
P.~Ghoshal$^1$,
S.~Gopinath$^1$, \\
J.~Gubeli$^1$, 
S.~Habet$^{1,2}$,
C.~Hernandez-Garcia$^1$,
A.~Hofler$^1$,
R.~Kazimi$^1$,
F.~Lin$^3$,
S.~Nagaitsev$^1$, \\
M.~Poelker$^1$,
B.~Rimmer$^1$,
Y.~Roblin$^1$,
V.~Lizarraga-Rubio$^4$,
A.~Seryi$^1$,
M.~Spata$^1$,
A.~Sy$^1$, \\
D.~Turner$^1$,
A.~Ushakov$^1$,
C.A.~Valerio-Lizarraga$^5$,
E. Voutier$^2$ \\
\it $^1$Thomas Jefferson National Accelerator Facility, Newport News, VA, USA \\
\it $^{2 }$Universit\'e Paris-Saclay, CNRS/IN2P3/IJCLab, France \\
\it $^{3 }$Oak Ridge National Laboratory, Oak Ridge, TN, USA \\
\it $^{4 }$Universidad de Guanajuato, Leon, Gto., M\'exico\\
\it $^{5 }$Universidad Aut\'onoma de Sinaloa, Culiac\'aan, M\'exico\\
}
 
\maketitle

\begin{abstract}
We present a scheme for the generation of a
high polarization positron beam with continous wave (CW) bunch structure for the Continuous Electron Beam Accelerator Facility (CEBAF) at Jefferson Laboratory (JLab).  The positrons are created in a high average power conversion target and collected by a CW capture linac and DC solenoid.
\end{abstract}

%\begin{abstract}
%We present a scheme for the generation of spin polarized positron beams with a continuous-wave (CW) bunch structure to be used at the Continuous Electron Beam Accelerator Facility (CEBAF) at Jefferson Laboratory (JLab).
%\end{abstract}

\section{Introduction}
The CEBAF accelerator has provided high energy spin polarized electron beams for almost 30 years.  Today, JLab is exploring an upgrade which would provide high energy spin polarized positron beams to address new physics~\cite{arrington2022physics, Accardi_2021}. 

A relatively new technique referred to as PEPPo (Polarized Electrons for Polarized Positrons) has been adopted~\cite{PhysRevLett.116.214801} to generate the positrons.  Here the spin polarization of an electron beam is transferred by polarized bremsstrahlung and polarized e+/e- pair creation within a high-power rotating tungsten target. 
%
%This technique was experimentally tested at %CEBAF previously \cite{PhysRevLett.116.214801} %and demonstrated efficiently the transfer of %the spin polarization from an 8.2 MeV/c %electron beam to positrons with polarization %approaching the electron polarization ( 85\%).  %Notably, this technique is essentially %independent of the initial electron beam %energy, providing great flexibility in the %choice of the electron injector to be used.   %Equally important, the bunch structure of the %incident electron beam is imprinted to the %positron beam, satisfying the critical %requirements.

\begin{figure}[!htb]
    \centering
    \includegraphics[width=77mm]{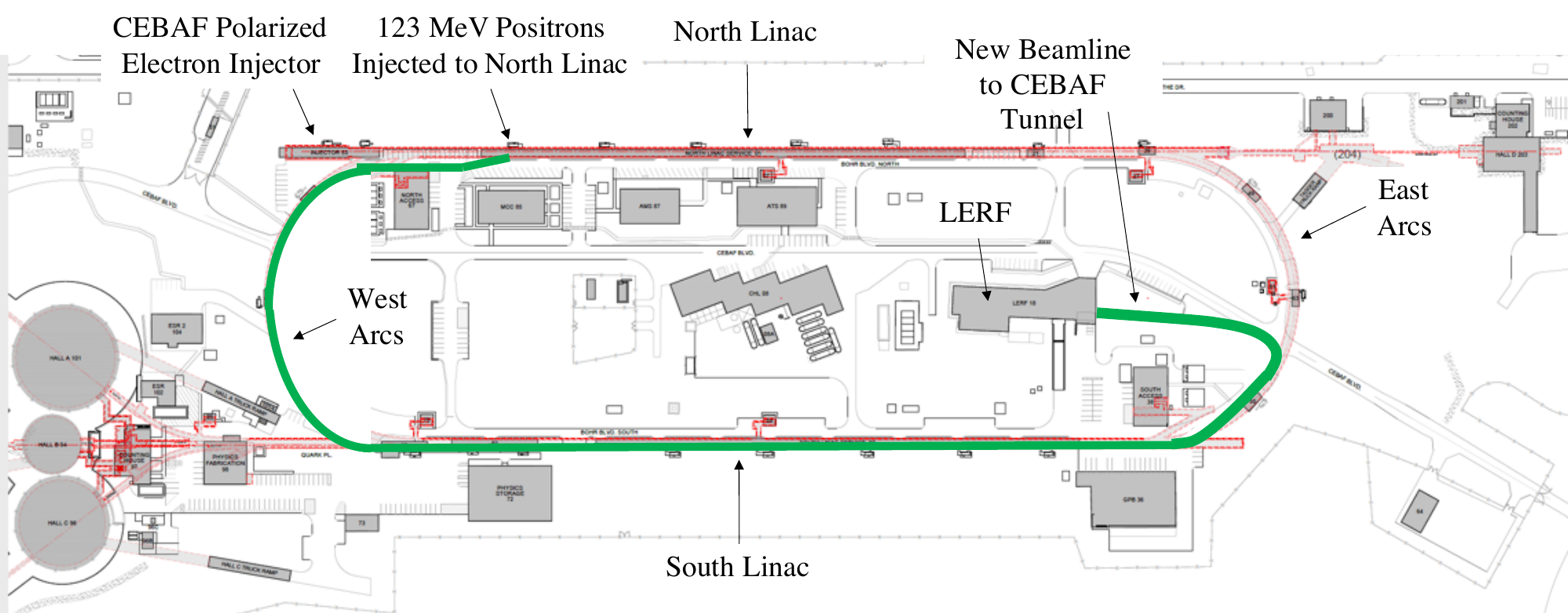}
    \caption{CEBAF and LERF accelerators.  Green line shows the new 123 MeV transport beam line connecting LERF to CEBAF for high energy acceleration of positron beams.}
    \label{fig:sl_to_nl}
\end{figure}

\begin{figure}[!htb]
    \centering
    \includegraphics[width=77mm]{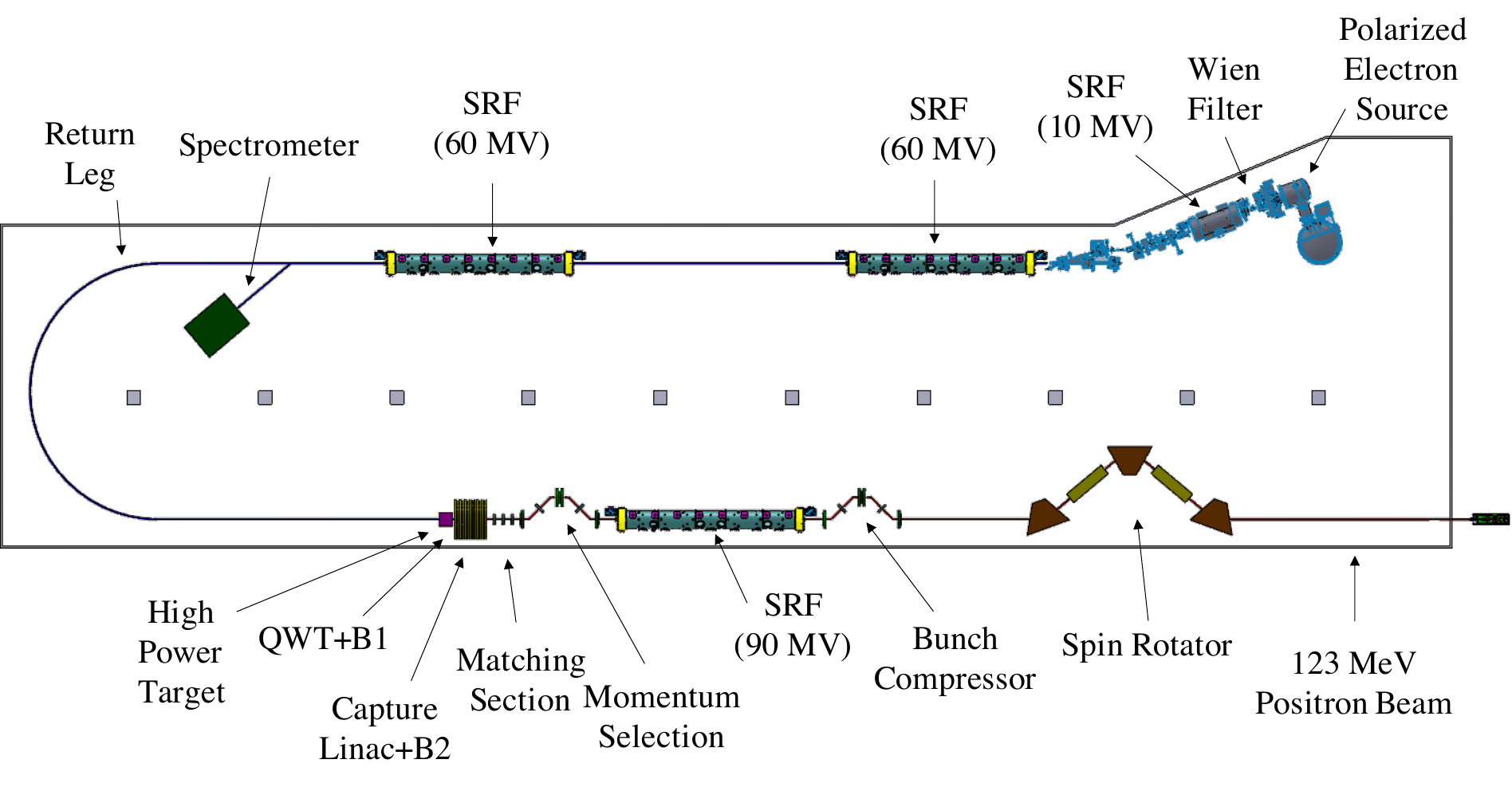}
    \caption{LERF layout of polarized e\textsuperscript{-} and e\textsuperscript{+} injectors.}
    \label{fig:lerf_layout}
\end{figure}

In this scheme two accelerators are used (see Fig.~\ref{fig:sl_to_nl}).   First, the Jefferson Lab Low Energy Recirculator Facility (LERF) building (see Fig.~\ref{fig:lerf_layout}) is repurposed to take advantage of existing electrical, cryogenic, and shielding facilities. A high current \SI{>1}{\milli\ampere} spin polarized CW electron beam is produced, accelerated to an energy of \SI{120}{\mega\electronvolt} and transported to the high-power target to generate the spin polarized positrons. Afterwards, the positrons are collected to maximize intensity or polarization, bunched and re-accelerated to \SI{123}{\mega\electronvolt}. Finally their spin direction may adjusted in a novel spin rotator.
 Once the positron beam exits the LERF it is transported from ground level through a new beam line to the CEBAF accelerator tunnel underground.  There it is transported half-way around the accelerator and injected as a usual electron beam would from the existing CEBAF electron injector.  The positrons are then accelerated to \SI{12}{\giga\electronvolt} and may be extracted at any pass (intermediate energies) to any of the four halls.
 The Ce\textsuperscript{+}BAF design is optimized to provide users with spin polarization \SI{>60}{\percent} at intensities \SI{>100}{\nano\ampere}, and with higher intensities when polarization is not needed.

\section{LERF}
\subsection{Polarized Electron Injector}
The existing LERF injector provides the baseline layout with the superconducting quarter cryomodule (Capture Linac: SRF 10 MV) capable of accelerating up to 10 mA CW beams to 9 MeV/$c$ ~\cite{FEL_Injector}. Upstream of the Capture Linac, the layout will resemble that in CEBAF, albeit more compact starting with the polarized electron source, followed by a Wien spin rotator and a buncher cavity for longitudinal matching to the SRF 10 MV. Downstream of the SRF 10 MV, a three dipole magnet chicane injects the electron beam into the first of two full-length accelerating SRF cryomodules (60 MV each).
The LERF electron gun will be a scaled-up version of the 130 keV inverted geometry gun used at CEBAF for many years~\cite{adderley2010load}. The CEBAF gun reliably provides highly spin polarized electron beams 90\% and average current of \SI{200}{\micro\ampere} with ~0.4 pC CW bunch trains (250/499 MHz). Due to excellent dynamic vacuum conditions and a biased anode limiting ionized residual gas from reaching the photocathode, charge lifetimes $>$400 C with strained-superlattice GaAs/GaAsP are achieved~\cite{adderley2022overview, yoskowitz2021improving}. 

However, because the bremsstrahlung yield of positrons from electrons will be low ($<10{^{-4}}$) a much higher beam current >1 mA is required with correspondingly higher bunch charge >2 pC.  We expect to operate the gun in a range of 300-350 kV to manage the higher bunch charge and allow direct injection to the SRF 10 MV. To meet the anticipated demands we expect to re-design the gun cathode electrode in two ways, (a) to have a larger spherical radius to achieve and safely maintain  higher gradient and (b) to accommodate a larger laser beam spot size to extend the charge lifetime.  Additionally the cathode must be free of field emission so we plan to include the capability of applying 50 kV beyond the required beam voltage for high voltage gas conditioning.

The higher bunch charges also pose challenges for the initial bunching and acceleration of the beam. Space charge forces will repel the electrons and reverse the bunching as the beam drifts, where space charge effects typically degrade beam quality.  To prevent this, we have kept the distance between the gun and the first accelerating element as short as possible and plan to compress the electron bunch from ~40 ps (determined by the optical pulsed) to about 2 ps within a few meters prior to the SRF QCM. 

The final section of the electron injector shapes the transverse emittance to match the acceptance for two CEBAF style CMs which accelerate the beam energy to about 120 MeV. A separate contribution to this conference~\cite{WEPA035} describes the electron injector in detail.

\subsection{High Power Target for Positron Production}
A conceptual design of the high power positron target has been developed. Tungsten has been chosen as the preferred target material. GEANT4~\cite{GEANT4:2002zbu} simulations have been used to determine that a tungsten target thickness of 4~mm is optimal for maximizing the Figure-of-Merit~\cite{habet:2023} (FOM, defined as the product of the positron current and the square of their longitudinal polarization). The thermal power deposited by a 1~mA electron beam current of 120~MeV energy into 4~mm of tungsten has been estimated with FLUKA~\cite{FLUKA1:ref,FLUKA2:ref} to be on the order of 17~kW. A typical target employed at JLab has less than 1~kW of electron beam power deposited into the target material. The only feasible cooling agent for the 17~kW target at Jefferson Lab would be water as the maximum cryogenic capacity for target cooling is less than 6~kW.  Notably, the only JLab target that surpassed the 1~kW mark to date was the 2.5~kW liquid hydrogen target~\cite{QWEAKTGT:ref} for the Qweak experiment. 

The design of the target has been done with ANSYS-Fluent~\cite{ANSYSCFD:ref} thermal simulations. Fluent calculations have shown that a static 4~mm thick tungsten target in a copper frame cooled by an internal water channel could sustain about 1~kW of electron beam power safely. A 35~cm diameter and 4 mm thick tungsten target rotating at 2~Hz could safely dissipate the 17~kW beam power deposited in it while maintaining a maximum temperature  below 1000~K. 
\begin{figure}[!htb]
    \centering
    \includegraphics[width=52mm]{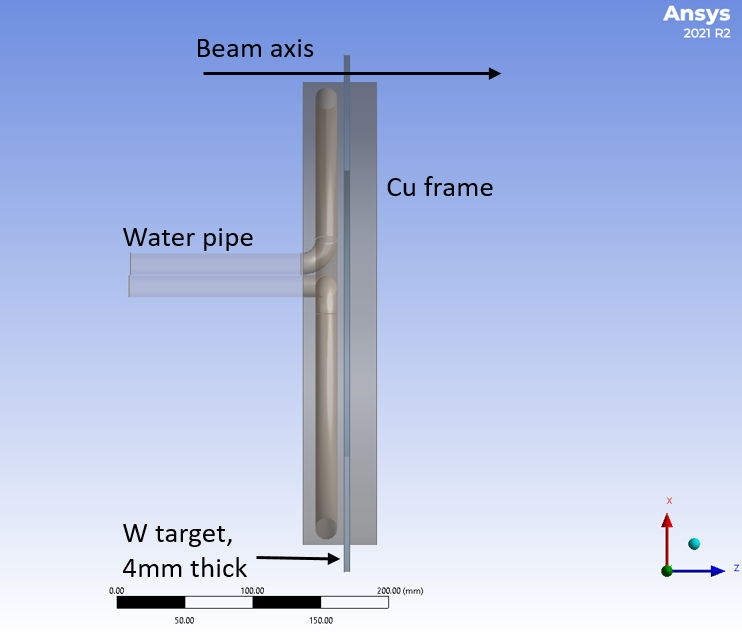}
    \caption{Side-view concept design of the rotating target.}
    \label{fig:rotatingtgt}
\end{figure}
A concept of the rotating tungsten target is shown in Fig.~\ref{fig:rotatingtgt}, where the tungsten is an annular ring partially encased in a copper frame. The electron beam impinges onto the 4~mm tungsten annulus a few mm from the frame and 17.5~cm from the rotation axis.  The conceptual design of the target will have to be engineered into a design that can be manufactured, operated and decommissioned safely. A detailed concept is evaluated in a separate contribution to this conference~\cite{WEPM120}.

\subsection{CW Positron Beam Formation}
The generation of positrons in a thick target creates an exceptionally broad distribution in transverse and longitudinal phase space.  A high field ($B_1$) quarter-wave transformer (QWT) located after the target decreases the transverse angular divergence of the positron distribution while also defining the central momentum of the positron polarization distribution to be collected. Following the high field region a low field ($B_2$) solenoid is used to manage the positron beam through an RF capture section. The solenoid fields $B_1$ and $B_2$ are optimized to maintain the large 4D transverse phase space of the beam~\cite{CHEHAB:} through this region.

A positron momentum spread $\delta p / p_0 < 1\%$ was chosen early on in the design in order to mitigate apertures in regions of large dispersion in the transport lines connecting the LERF to CEBAF. Although we have not settled on a final momentum spread this issue motivated us to include an RF capture section right after the QWT in our design studies in order to decrease the longitudinal energy spread as well as improve the transverse beam emittance.

Following the RF capture region the positron momenta is defined by a chicane beamline composed of quadrupoles and dipoles to create a correlation between positron energy and transverse position at its midpoint.  After the chicane the positron beam is accelerated in a  SRF CM to 123~MeV (an injection energy requirement for CEBAF 12 GeV) and transported through a bunch compression chicane to achieve a bunch length of a few picoseconds.

In this contribution the chicane was optimized for 60~MeV/$c$ positrons (maximum polarized FoM) while passing a 1\% energy spread.   Results of recent CW polarized positron beam simulations are shown in Table~\ref{ce+bafparam} relative to our present design goals.  We have met or exceeded all goals except for the normalized emittance, which can be met by reducing the acceptance and reducing the positron beam current or increasing the drive beam power.  A reference to earlier work on this topic was reported in Ref.~\cite{Habet:2022fch}. 
\begin{table}
\caption{ Simulated parameters of the Ce\textsuperscript{+}BAF injector.}
\begin{center}
\begin{tabular}{lrrr}
\toprule
\textbf{Ce$^+$BAF Parameter} & \textbf{Status} &  \textbf{Goal}  \\
\midrule
$p_0$ [MeV/$c$] & 60 & 60 \\
$\sigma_{\delta p/p_0}$ [\%] &    \color{black}{0.68} &  $\pm$1                \\
$\sigma_{z}$ [ps] &    \color{black}{3} &              $\leq 4$    \\     
Normalized $\epsilon_{n} $ [mm~mrad]  &  \color{black}{140}                & $\leq 40$ \\
 $p_f$ [MeV/$c$] & \color{black}{123} & 123 \\
 $I_{\mathrm{e}^+} (P>60\mathrm{\%})$ [nA] & \color{black}{170} & > 50 \\  
\bottomrule
\end{tabular}
\end{center}
\label{ce+bafparam}
\end{table}
\subsection{Positron Spin Rotator}
The precession of the electron beam polarization when accelerated at CEBAF to 12 GeV is more than 60 full revolutions.  Experiments however most often require longitudinal or sometimes transverse spin polarization at their target.  At CEBAF a 4$\pi$ spin rotator consisting of two Wien filters with intervening solenoid magnets~\cite{WF_Grames} is used to orient the spin at the injector to control the final spin polarization at the experiment.  This is convenient when the beam energy is ~100 keV and the required Wien filter field strengths are modest (e.g. E ~1 MV/m and B ~100 G).  However, the positron beam production energies at the LERF are 10's of MeV and the final beam energy is >100 MeV, making a Wien filter impractical.

\begin{figure}[!htb]
    \centering
    \includegraphics[width=70mm]{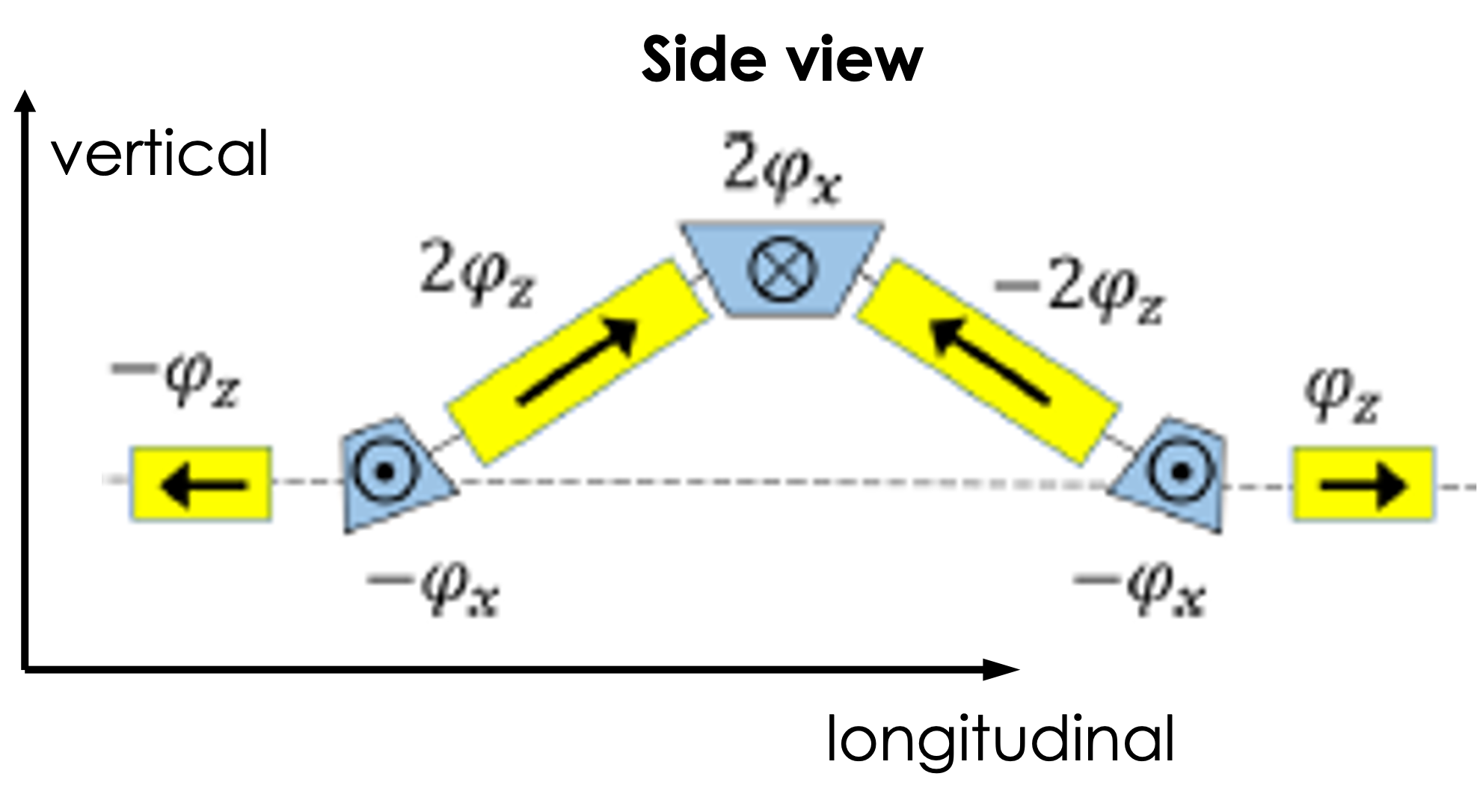}
    \caption{Spin rotator concept. $\phi_x$: spin rotation around the radial axis, $\phi_z$: spin rotation around the longitudinal axis.}
    \label{fig:spin_rotator}
\end{figure}

For Ce\textsuperscript{+}BAF a  higher energy spin rotator concept has been imagined.  The proposed spin rotator scheme is shown in Fig.~\ref{fig:spin_rotator}.  Composed by interleaved dipole and solenoid fields the small anomalous gyromagnetic factor for positrons (or even electrons) means the spin rotation in the solenoids is more effective than in the dipoles at lower energies~\cite{SR_Filatov}. However, the dipole magnetic field is necessary to provide the desired spin rotation axis. Rotating the spin around the longitudinal solenoid and radial dipole fields, this spin rotator can provide a desired net spin rotation around the vertical axis in the horizontal plane. Notably in this design the dipole fields are arranged with net zero bending angle, leaving the beam trajectory intact and transparent to beam orbit perturbations.  Further details of this design will be presented in a future presentation once on-going simulations are completed. 

\section{\NoCaseChange{12 GeV Ce\textsuperscript{+}BAF}}
Once the CW positron beam has been formed and the spin oriented it is ready for acceleration to higher energies.
%has been generated, captured and manipulated to define the transverse emittance and longitudinal parameters, it is accelerated to 123 MeV/c  and directed to the south side of the LERF via a beamline designed to accommodate for the large energy spread and transverse emittance by using a double bend achromat (DBA) cell and short dipoles \cite{DBA-CAI}. It is followed by a long straight FODO channel section optimized for maximum acceptance (Figure~\ref{fig:lerf_layout}). A spin rotator is located in that long straight section.
%
%\begin{figure}[!htb]
%    \centering
%    \includegraphics[width=77mm]{Figures/lerf_layout.png}
%    \caption{Layout of the positron beamline in the LERF vault.}
%    \label{fig:lerf_layout}
%\end{figure}
%
 The positron beam is transported in a new tunnel connecting the east side of LERF to the south east corner of CEBAF near the entrance of the South Linac (Fig.~\ref{fig:sl_to_nl}). This new beamline features a double-bend achromat (DBA) to maintain small dispersion and a vertical achromatic translator to bring the beam to the elevation of the CEBAF South Linac tunnel near the ceiling. 
 %
 %\begin{figure}[!htb]
 %   \centering
 %   \includegraphics[width=77mm]{Figures/lerf_to_ea.png}
 %   \caption{Tunnel and beamline connecting the LERF to the east side of CEBAF.}
 %   \label{fig:lerf_to_ea}
%\end{figure}
%
At this point a long FODO channel attached to the ceiling of the South Linac transports the beam to the west side of CEBAF where it is bent 180 degrees via a DBA-like lattice with low dispersion and is also isochronous. At the end of this long transport line a vertical achromat translator and horizontal bending magnets bring the beam to the start of the North Linac where it is injected.  Additionally, each beamline also has a betatron matching section.

While this long beamline from LERF to CEBAF is designed for the 123 MeV/$c$ positron beam it should also be suitable for an electron beam with energy up to 650 MeV/$c$ to be compatible with a future upgrade of CEBAF to 22 GeV.

 The CEBAF accelerator limits the maximum transverse emittance that one can transport because of the reduced acceptance at the extraction corners. We estimate that one can inject between 40 and 120 mm.mrad of normalized emittance at the front of the north linac. In terms of longitudinal acceptance, we are planning to change the optics configuration for the first two recirculation arcs (east and west sides) in order to have smaller dispersion functions and an easily tunable momentum compaction. With these new optics we should expect to inject up to a percent of energy spread in the front of the north linac and transport a beam that has a longitudinal bunch length around 1 mm.   A separate contribution to this conference~\cite{MOPM081} is exploring the admittance of the electron injector and first recirculation pass of CEBAF.
 
\section{OUTLOOK}
The Ce\textsuperscript{+}BAF working group has developed a scheme to provide CEBAF with polarized positron beams with CW time structure. Early designs and simulated parameters combined with constraints are approaching the anticipated goals.  Our focus in the coming months is to develop a white paper documenting in greater detail the technical approach and additional issues being addressed, but not reported in the length of these proceedings.

%One of the main challenges ahead is refining our scheme into a more realistic configuration of both the beam line components (e.g. high field magnets and high current photoguns) as well as managing the high beam power (e.g. the target and radiation shielding).

%
% only for "biblatex"
%
\ifboolexpr{bool{jacowbiblatex}}%
	{\printbibliography}%
	{%
	% "biblatex" is not used, go the "manual" way

} % end \ifboolexpr

%
% for use as JACoW template the inclusion of the ANNEX parts have been commented out
% to generate the complete documentation please remove the "%" of the next two commands
% 
%%%\newpage

%%%\include{annexes-A4}

\end{document}